\documentclass{statsoc}

\usepackage[a4paper]{geometry}
\usepackage{graphicx}
\usepackage[textwidth=8em,textsize=small]{todonotes}
\usepackage{amsmath}
\usepackage{natbib}
\usepackage{macros}
\usepackage{centernot}
\usepackage{rotating}
\usepackage[colorlinks,citecolor=blue,urlcolor=blue]{hyperref}

\usepackage{setspace}
\newcommand{\beginsupplement}{%
        \setcounter{table}{0}
        \renewcommand{\thetable}{S\arabic{table}}%
        \setcounter{figure}{0}
        \renewcommand{\thefigure}{S\arabic{figure}}%
     }

\title[Two Sample Test for Eigendecompositions of Functional Data]{Two Sample Test for Eigendecompositions of Functional Data}
\author[Author 1 {\it et al.}]{\'Angel Garc\'ia de la Garza}
\address{Division of Biostatistics, Albert Einstein College of Medicine,
New York City, 
USA.}
\email{angel.garciadelagarza@einsteinmed.edu}
\author[Author 2]{Britton Sauerbrei}
\address{Department of Neurosciences, Case Western Reserve University,
Cleveland, 
USA.}
\author[Garc\'ia de la Garza et al.]{Jeff Goldsmith}
\address{Department of Biostatistics, Columbia University ,
New York City, 
USA.}

\begin{document}
\begin{abstract}
Neuron-level firing data is believed to be governed by latent activation patterns during task completion. Analysing repeated trials of a task allows us to study these patterns, typically by averaging in-vivo neural spikes across trials. However, estimates of underlying latent activation patterns show trial-to-trial variability. Our aim is to determine whether this variation arises from observed data differences or changes in the latent activation patterns themselves. The latter would imply that current approaches overlook meaningful activation changes, necessitating adjustments in dimension reduction and downstream analysis.  We propose a test that compares the eigendecompositions of two samples of functional data based on the covariance matrix of scores derived from a functional principal component analysis of the pooled data. Initially developed for independent samples, we later extend the test to paired samples, as necessary for our data. Simulation studies demonstrate its superior power compared to leading methods across various scenarios. In an experiment with 157 trials, we analyse all pairwise comparisons using a permutation approach to test the null hypothesis of shared latent activation patterns across trials. Our findings reveal trial-to-trial variation in latent activation patterns that cannot be attributed to sampling noise.


\end{abstract}

\newpage

\section{Introduction}

In-vivo cell-specific neural spike measurements from animals have been fundamental in understanding the relationship between the brain and behaviour \citep{brown2004multiple, briggman2005optical, gerstner2014neuronal,sauerbrei_cortical_2020}. The observed spike behaviour of an individual neuron is typically thought to be a realisation of an underlying latent neural process that describes or generates activation across populations of neurons  \citep{yu_gaussian-process_2008}. This motivates an interest in estimation of activation patterns, and indeed activation patterns reconstructed from raw neural spikes have been shown to better explain behaviour compared to using only neuron-level observations \citep{broome2006encoding,santhanam2009factor}. There is evidence that activation patterns can change in response to new experimental conditions \citep{churchland2010stimulus, shenoy2013cortical, hartmann2015s}, but also that they can remain constant when experimental conditions are unchanging, as when an animal repeatedly performs the same task \citep{flint2016long,gallego2020long}. Assumptions about data generating processes can affect analysis decisions: if it is assumed that activation patterns are unchanging they might be estimated by aggregating across repeated trials of the same experimental condition. 

\begin{figure}[htb]
  \centering
     \begin{tabular}{cc}
       \includegraphics[width= 0.9\textwidth]{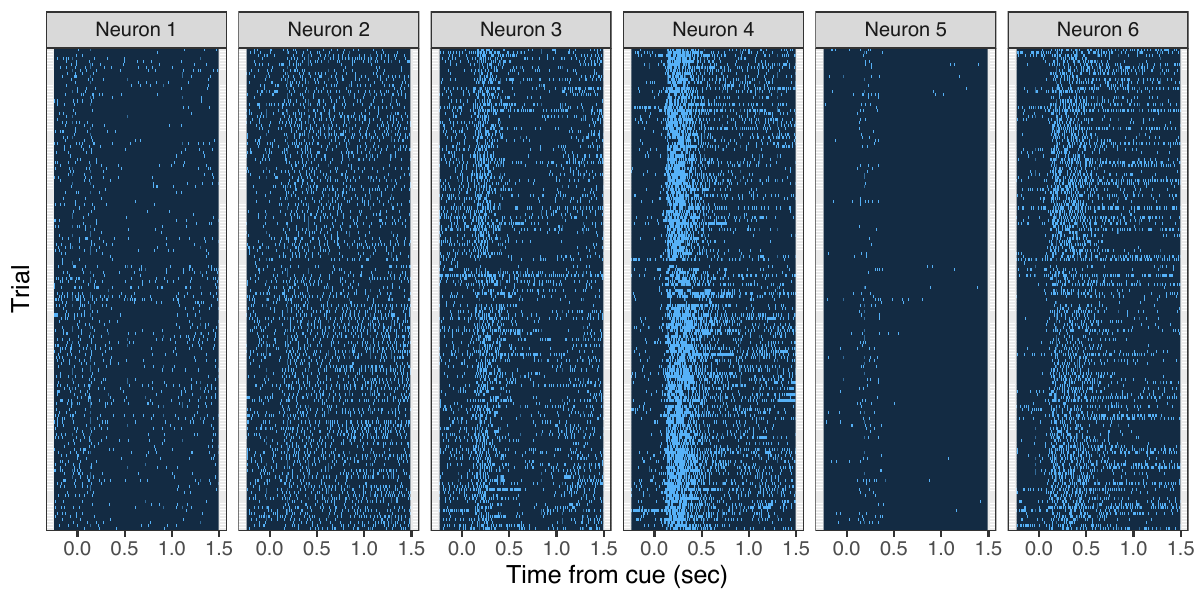} 
     \end{tabular}
     \caption{Panel A displays a lasagna plot of the activation of six example neurons across 174 timepoints and 157 trials. Light blue indicates that the neuron is activate at that specific instance.}
    \label{fig:raw_data}
\end{figure}

Our goal is to determine whether activation patterns in the motor cortex remain constant across trials in an experiment where a trained mouse reached for a food pellet after hearing an auditory cue \citep{sauerbrei_cortical_2020}. We analyse spike data from 25 neurons in the mouse's motor cortex recorded during 157 trials, each lasting 1.75 seconds. Our measurements begin 0.25 seconds before the auditory cue and were recorded in 10ms intervals aligned with the cue. A subset of the resulting data are shown in Figure \ref{fig:raw_data}, which depicts the activation patterns of six representative neurons and reveals commonalities in their activation. Specifically, neural activity generally increases immediately following the cue and gradually decreases back to the pre-cue baseline level. The magnitude of the activation increase and the time to return to baseline is not constant across neurons, however, and one neuron decreases in activation following the cue. These shared patterns may suggest that each neuron spikes under a common low-dimensional latent activation process or data-generating mechanism. The variation in measurements from trial to trial may be due to changes in the mechanism between trials, which could be related to subtle differences in the voluntary reaching movement.

We will use Functional Principal Component Analysis (FPCA) \citep{yao2005functional, goldsmith2013corrected,xiao2016fast} to model the trial-specific latent activation patterns that describe observed spike data at the neuron level. We will then test the hypothesis that the activation patterns obtained through FPCA are the same across all trials; to do this, we develop a novel two-sample test that compares activation patterns in terms of FPCA scores from a dataset constructed by pooling the two samples. Because data in this experiment are observed on the same neurons, we also develop a paired version of this test. Next, we will make all possible pairwise comparisons of activation patterns across trials using this paired test, and summarise the results by comparing the distribution of obtained p-values to a uniform distribution. Finally, we will generate a permutation-based null distribution of this summary statistic to test the global hypothesis that activation patterns are the same across all trials. To our knowledge, this is the first formal examination of whether observed differences in the estimated activation patterns are attributable to differences in the true latent activation process.

We now briefly review existing tests to compare two samples of functional data. Given our contributions, we emphasize second-order moment tests for differences in eigendecompositions or, equivalently, covariance operators. Examples of such tests include the eigenfunction-based approximation of the distance between two covariance operators proposed by \citet{panaretos_second-order_2010}, which assumed Gaussian random functions and two independent groups. Extensions to non-Gaussian data have been proposed by \citet{fremdt_testing_2013}, to tests of more than two groups by \citet{boente2018testing}, and to dependent data by \citet{zhang2015two}. Additional second-order tests include the Procrustes distance test developed by \citet{pigoli2014distances}, and the M-test that is robust to non-Gaussian observations developed by \citet{kraus2012dispersion}. Several two-sample tests for functional data compare the centre of the two groups; examples include the point-wise t-test for differences in functional means proposed by \citet{ramsay_functional_2005}, $L^2$ norm-based tests \citep{zhang_two_2010, horvath2012inference, benko_common_2009, zhang_two_2010, paparoditis_bootstrap-based_2016, kashlak_analytic_2022} or the likelihood-ratio approach proposed by \citet{staicu2014likelihood}. Finally, tests to compare the distribution of the two samples have been proposed by \citet{benko_common_2009} and \citet{pomann_two-sample_2016}, among others. These jointly test the mean function, eigenvalues, and eigenvectors from an FPCA decomposition, and use an Anderson-Darling Rank test on FPC scores, respectively.

Our proposed test uses the covariance matrices of scores from an FPCA decomposition, so we will also briefly review the extensive literature on comparing two covariance matrices in multivariate non-functional settings. Examples for low-dimensional data include likelihood ratio tests \citep{anderson_introduction_1962, sugiura1968unbiasedness, gupta1973properties,o1992robust} and sampling-methods tests \citep{zhang1992bootstrap,zhu2002resampling,yang2012resampling}. For high-dimensional data (i.e., $p > n$), spectral analysis approaches \citep{yin1988limit, bai1988limiting, zheng2017clt} and Frobenius norm-based tests \citep{schott_test_2007, srivastava_testing_2010, li2012two} have been developed. We base our testing procedure on the work of \citet{cai_two-sample_2013}, which proposed the standardised maximum difference between the entries of two covariance matrices as a test statistic. This test has several advantages: it makes no distributional assumptions, has an analytic limiting distribution, and has significantly higher power than other methods in simulation studies. 

The rest of this manuscript is structured as follows. Section 2 introduces our two-sample test for the equality of FPCA decompositions using the covariance matrix of FPCA scores derived from a dataset that pools the samples. We extend this framework to paired data, a setting that is necessary for our application but for which limited literature exists. In Section 3, we present simulation studies designed to evaluate the performance of our proposed test and compare to existing methods. We apply our methodology to the neural spike data and discuss the scientific implications in Section 4. Section 5 contains conclusions and discusses possible future avenues for research.

%%%%%%%%%%%%%%%%%%%%%%%%%%%%%%%%%%%%%%%
%%%%%%%%%%%%%%%%%%%%%%%%%%%%%%%%%%%%%%%
%%%%%%%%%%%%%%%%%%%%%%%%%%%%%%%%%%%%%%%

\section{Methods} 
\label{sec:methods}

Let the two sets of curves $\left[Y_{i}^{(1)}(t) : i \in \left\{1, \dots, I_{1}\right\}\right]$ and $\left[Y_{i}^{(2)}(t) : i \in \left\{1, \dots, I_{2}\right\}\right]$ be realisations from two random processes, $Y^{(1)}(t)$ and $Y^{(2)}(t)$, over $t \in [0,1]$. Further assume that these random processes have corresponding mean functions $\mu^{(1)}(t)$ and $\mu^{(2)}(t)$, and covariance operators $\Sigma^{(1)}(s,t)$ and $\Sigma^{(2)}(s,t)$. Using Mercer's theorem, let $\Sigma^{(z)}(s,t) \approx \sum_{k=1}^{K} \left[\lambda_{k}^{(z)} \phi_{k}^{(z)}(s) \phi_{k}^{(z)}(t)\right]$ be the $K$-truncated spectral decomposition of $\Sigma^{(z)}(s,t)$ for group indicator $z \in \{1,2\}$, where $\boldsymbol{\Phi}^{(z)}(t) = \left\{\phi_{k}^{(z)}(t): k \in \{1,\dots,K\}\right\}$ are orthonormal eigenfunctions and $\boldsymbol{\lambda}^{(z)} = \left\{\lambda_{k}^{(z)}: k \in \{1,\dots,K\}\right\}$ are the corresponding eigenvalues.

Our scientific goal is to test the hypothesis that the shape $\boldsymbol{\Phi}^{(z)}(t)$ and scale $\boldsymbol{\lambda}^{(z)}$ of the eigenfunctions obtained from two functional datasets is the same. We can frame this test equivalently using the covariance operator defined through the spectral decomposition of $\Sigma^{(z)}(s,t)$. That is, we focus on using two samples of data to test the hypothesis 
\begin{equation}
\label{eq:hyp_test}
  H_0: \Sigma^{(1)}(s,t) = \Sigma^{(2)}(s,t) \quad \mbox{versus} \quad H_1: \Sigma^{(1)}(s,t) \neq \Sigma^{(2)}(s,t)
.\end{equation}
We present methodology to test (\ref{eq:hyp_test}) for two independent samples of functions in Section \ref{subsec:testing_procedure}, extend this procedure to paired functions in Section \ref{subsec:testing_pairs}, and describe the implementation for real-data settings in Section \ref{subsec:implementation}. 

%%%%%%%%%%%%%%%%%%%%%%%%%%%%%%%%%%%%%%%
\subsection{Testing Procedure for Independent Realisations}
\label{subsec:testing_procedure}

Our approach for testing (\ref{eq:hyp_test}) begins by pooling the data from the two samples, then performs FPCA on the pooled sample, and finally compares the covariance of the scores for each group derived from this pooled FPCA analysis. Although the specific implementation and goals of our test differ from those in \cite{pomann_two-sample_2016}, we note that our use of a pooled dataset for FPCA is related to the ``marginal" covariance in that work's two-sample test. 

Assume that curves $\left[Y_{i}^{(1)}(t) : i \in \left\{1, \dots, I_{1}\right\}\right]$ and $\left[Y_{i}^{(2)}(t) : i \in \left\{1, \dots, I_{2}\right\}\right]$ are independent realisations of their respective data generating mechanisms. Further, we assume that the data for the two groups are de-trended, or equivalently $\mu^{(1)}(t)=\mu^{(2)}(t)=0$. Define the pooled sample $X(t)=\left[Y_{i}^{(z)}(t) : i \in \left\{1, \dots, I_{z}\right\}, z \in \left\{1, 2\right\}\right]$, and let $\Sigma(s,t) = \text{cov}\left\{X(s),X(t)\right\}$ be the pooled (i.e. marginal over $z$) covariance function of $X(t)$. Using Mercer's theorem, we express the $K$-truncated spectral decomposition of the pooled covariance as $\Sigma(s,t) \approx \sum_{k=1}^{K} \left[\lambda_{k} \phi_{k}(s) \phi_{k}(t)\right]$ where $\boldsymbol{\Phi}(t) = \left\{\phi_{k}(t): k \in \left\{1,\dots,K\right\}\right\}$ is the set of pooled eigenfunctions and $\boldsymbol{\lambda} = \left\{\lambda_{k}: k \in \left\{1, \dots, K \right\} \right\}$ is the set of pooled eigenvalues. Finally, we define the $K$-dimensional approximation of $Y^{(z)}_{i}(t)$ to be:
\begin{equation}
\label{eq:null_model}
  Y^{(z)}_{i}(t)\approx \sum_{k=1}^{K}\left[\zeta^{(z)}_{ik}\phi_k(t)\right] 
\end{equation}
where the $i^{th}$ observation in group $z$ is decomposed into, the pooled set of eigenfunctions $\boldsymbol{\Phi}(t)$, and the curve-specific scores $\zeta_{ik}^{(z)}={\displaystyle \int_{0}^{1} } \big\{ \big[Y_i^{(z)}(t)-\mu^{(z)}(t)\big]\phi_{k}(t)\big\} dt$, assumed to be uncorrelated random variables with mean zero and variance $\lambda_k$. We now describe a test for (\ref{eq:hyp_test}) using the covariance matrix of the scores coming from the preceding model.

\begin{prop}

Define $\boldsymbol{\zeta}_i^{(z)} = \left\{\zeta_{ik}^{(z)}: k \in \{1,\dots,K\}\right\}$ to be the $K \times 1$ random vector of scores derived from model \eqref{eq:null_model} for subject $i \in \left\{1,\dots,I_{z}\right\}$ in group $z \in \{1,2\}$ with covariance matrix $\boldsymbol{\Omega}^{(z)}$. Then, the hypothesis test \eqref{eq:hyp_test} is equivalent to 

\begin{equation}
\label{eq:hyp_test_sigma}
  H_0: \boldsymbol{\Omega}^{(1)} = \boldsymbol{\Omega}^{(2)} \quad \mbox{versus} \quad H_1: \boldsymbol{\Omega}^{(1)} \neq \boldsymbol{\Omega}^{(2)}.
\end{equation}
\end{prop}

To see that the proposition holds, note that the covariance operator for any function $Y^{(z)}_i(t)$ under model \eqref{eq:null_model}, which assumes a shared eigendecomposition, is
\begin{equation}
\begin{aligned}
\label{eq:cov_operator_null_model}
      \text{cov}\!\left[Y^{(z)}_i(t), Y^{(z)}_i(s)\right] &= \sum_{p = 1}^K \sum_{q = 1}^K \left[ \text{cov}\!\left(\zeta_{ip}^{(z)}, \zeta_{iq}^{(z)}\right) \phi_p(t) \phi_q(s) \right]  \quad z \in \{1,2\}
\end{aligned}           
\end{equation}
Hence, for model (\ref{eq:null_model}), $\Sigma^{(1)}(s,t) = \Sigma^{(2)}(s,t)$, when $\boldsymbol{\Omega}^{(1)} = \boldsymbol{\Omega}^{(2)}$ and vice-versa. Consequently, we test the null hypothesis (\ref{eq:hyp_test}) through the score covariance matrices derived under model \eqref{eq:null_model}. 

\begin{figure}[htbp]
  \centering
     \begin{tabular}{cc}
       \includegraphics[width= \textwidth]{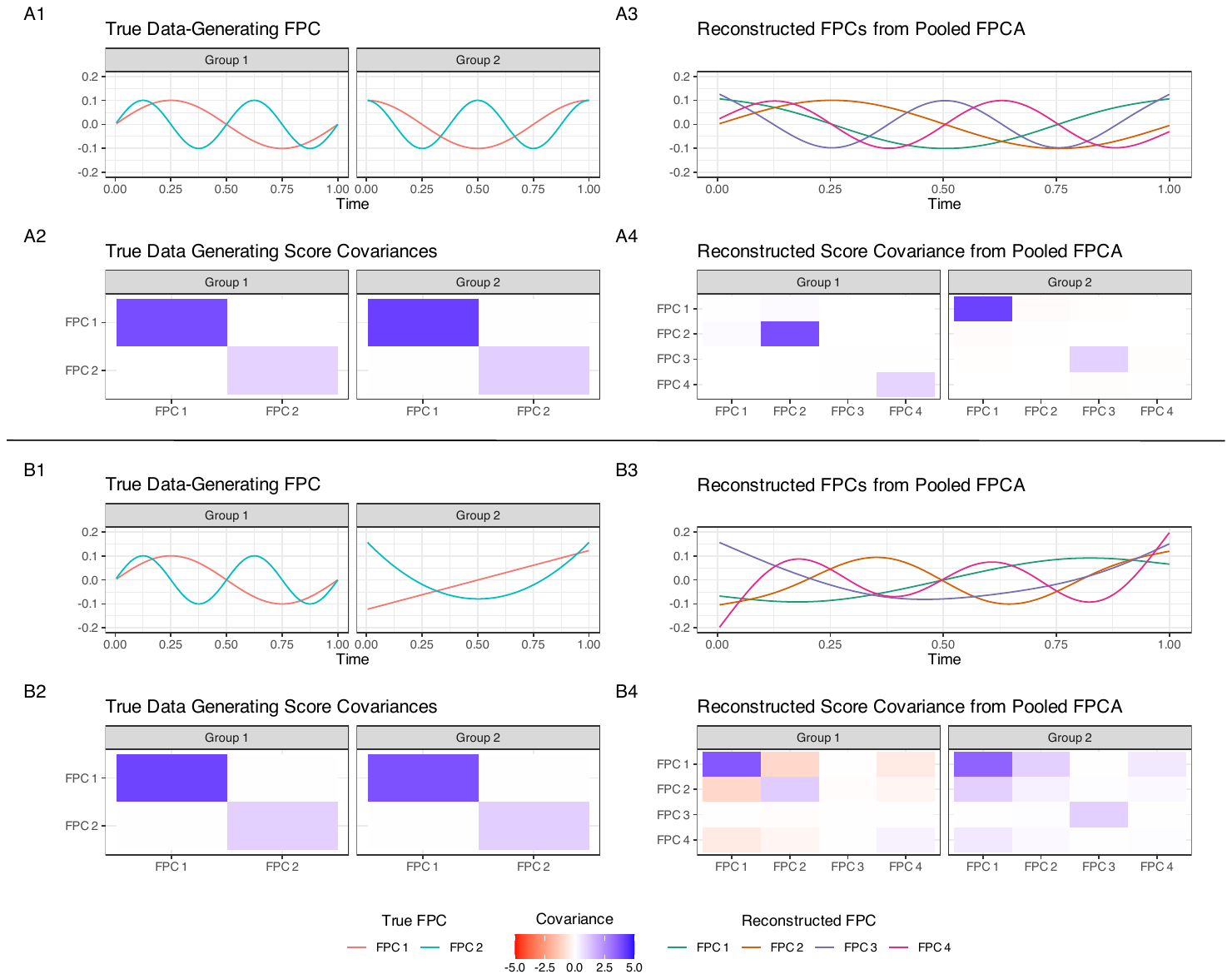} 
     \end{tabular}
     \caption{Panel A1-4 displays scenarios in which the FPCs across groups are orthogonal. Panels B1-4 show data simulations in which the FPCs across groups are not orthogonal. Panels A1 and B1 depict the true data-generating FPCs used in the simulations. Panel A2 and B2 display the true score covariance matrix used to generate the data. Panels A3 and B3 show the reconstructed FPCs from a pooled FPCA. Panels A4 and B4 demonstrate the score covariance matrix obtained from the pooled FPCA decomposition.}
    \label{fig:method_plot}
\end{figure}

We present a schematic of our method in Figure \ref{fig:method_plot}. We generate data for two groups with distinct covariance operators in two different scenarios. In the top two rows, the FPCs across the two groups are orthogonal while the within-group score covariances are identical. In these rows, the left panels show the true FPCs and score covariances that generate data in the two groups. The right panels, meanwhile, show the FPCs from a pooled analysis, along with the corresponding score covariance matrices $\boldsymbol{\Omega}^{(1)}$ and $\boldsymbol{\Omega}^{(2)}$ derived under (\ref{eq:null_model}). As expected, the FPCs from the pooled dataset combine the (orthogonal) FPCs that generate data in each sample, and the covariances are partitioned accordingly. The bottom two rows in Figure~\ref{fig:method_plot} show an example in which FPCs across groups are not orthogonal. In this example, the FPCs derived from the pooled sample are combinations of those in the two groups, and the score covariances are non-diagonal matrices. This Figure emphasises that the score covariance matrices $\boldsymbol{\Omega}^{(1)}$ and $\boldsymbol{\Omega}^{(2)}$ can be used to distinguish between data generating mechanisms, including settings where the un-pooled samples have the same score covariances. Additional details of data generation in each simulation scenario can be found in the supporting web materials.

We test for equality of the score covariance matrices $\boldsymbol{\Omega}^{(1)}$ and $\boldsymbol{\Omega}^{(2)}$ using the standardised maximum difference \citep{cai_two-sample_2013}. Intuitively, two covariances will be different if and only if the maximum absolute difference in the same entry of the two covariances is greater than zero. More concretely, we test $H_0: \boldsymbol{\Omega}^{(1)} = \boldsymbol{\Omega}^{(2)}$ through the hypothesis $H_0: \text{max}_{1 \leq p \leq q \leq K} \left|\omega_{pq}^{(1)} - \omega_{pq}^{(2)}\right| = 0$, where $\omega_{pq}^{(z)}$ is the $pq^{th}$ entry of $\boldsymbol{\Omega}^{(z)}$. To define a practical test statistic, let $\hat{\omega}_{pq}^{(z)}$ be the $pq^{th}$ entry of score sample covariance matrices in group $z$:
\label{eq:sample_covariance}
\begin{align}
\begin{split}
    \hat{\boldsymbol{\Omega}}^{(z)} &= \frac{1}{I_z} \sum_{i=1}^{I_z}\left[\left(\boldsymbol{\zeta}^{(z)}_{i} - \bar{\boldsymbol{\zeta}}^{(z)}\right)\left(\boldsymbol{\zeta}^{(z)}_{i} - \bar{\boldsymbol{\zeta}}^{(z)}\right)^T\right]  \\  \bar{\boldsymbol{\zeta}}^{(z)} &= \frac{1}{I_z} \sum_{i=1}^{I_z} \left[\boldsymbol{\zeta}^{(z)}_{i}\right], \quad z \in \{1,2\}. 
\end{split}
\end{align}

The test statistic $M$ is the maximum standardised difference in any entry of the sample covariance matrices:
\begin{equation}
    \label{eq:test_statistic}
    M = \text{max}_{1 \leq p \leq q \leq K} \frac{\left[\hat{\omega}_{pq}^{(1)} - \hat{\omega}_{pq}^{(2)}\right]^{2}}{\left[\hat{\theta}_{pq}^{(1)}/I_1 + \hat{\theta}_{pq}^{(2)}/I_2\right]}.
\end{equation}
The standardisation term in the denominator, defined as 
\begin{equation}
\label{eq:sample_variance_covariance}
\begin{split}
    \hat{\theta}_{pq}^{(z)} &= \frac{1}{I_z} \sum_{i=1}^{I_z} 
    \left[\left(\zeta_{ip}^{(z)} - \bar{\zeta}_{p}^{(z)}\right)\left(\zeta_{iq}^{(z)} - \bar{\zeta}_{q}^{(z)}\right) - \hat{\omega}_{pq}^{(z)}\right]^{2}, \quad z \in \{1,2\} \\
\end{split}
\end{equation}
where $\bar{\zeta}_{p}^{(z)}$ is the $p^{th}$ entry of the vector $\bar{\zeta}^{(z)}$, accounts for heterogeneity in the estimates of the difference in each entry of the sample covariance matrices. 

When $H_0: \boldsymbol{\Omega}^{(1)} = \boldsymbol{\Omega}^{(2)}$ holds, the value $M - 4\log K + \log \log K$ converges to a Gumbel extreme distribution, and a closed form for the asymptotic rejection threshold exists. \citet{cai_two-sample_2013} proposed an adjustment to the critical value of the test statistic for small sample sizes by using simulated standard Gaussian datasets to estimate the distribution of the test statistic under the null. We address the case of small sample sizes through a permutation-based empirical null distribution of the test statistic.

%%%%%%%%%%%%%%%%%%%%%%%%%%%%%%%%%%%%%%%

\subsection{Testing Procedure for Dependent or Paired Realisations}
\label{subsec:testing_pairs}

The test for covariance operators developed in Section \ref{subsec:testing_procedure}  compares the covariance matrices of scores obtained from FPCA using a test that assumes independent samples. Because our motivating data consists of repeated observations on the same study units, we now extend that test to paired data. 

Consider a sample of paired curves $\left\{\left[Y_{i}^{(1)}(t),Y_{i}^{(2)}(t)\right] : i \in \left\{1, \dots, I\right\}\right\}$ that are realisations of the random processes $Y^{(1)}(\cdot)$ and $Y^{(2)}(\cdot)$. Define the sets of subject-specific scores $\left\{\boldsymbol{\zeta}^{(1)}_{i}: i \in \{1,\dots,I\}\right\}$ and $\left\{\boldsymbol{\zeta}^{(2)}_{i}: i \in \{1,\dots,I\}\right\}$ with covariances $\boldsymbol{\Omega}^{(1)}$ and $\boldsymbol{\Omega}^{(2)}$ coming from the model that assumes a pooled eigendecomposition (\ref{eq:null_model}). Although these scores are obtained without modelling the paired data structure, we can account for the pairing by including a term for the correlations between scores when standardising our test statistic. We define the updated test statistic for comparing eigendecompositions of paired functions as:
\begin{equation}
\label{eq:pair_teststat}
    M = \text{max}_{1 \leq p \leq q \leq K} \frac{\left[\hat{\omega}_{pq}^{(1)} - \hat{\omega}_{pq}^{(2)} \right]^{2}}{\left[\left(\hat{\theta}_{pq}^{(1)} + \hat{\theta}_{pq}^{(2)} -2\hat{\phi}_{pq}\right)/I\right]}
\end{equation}
where $\hat{\omega}_{pq}^{(1)}$, $\hat{\omega}_{pq}^{(2)}$ are defined as in (\ref{eq:sample_covariance}), $\hat{\theta}_{pq}^{(1)}$, $\hat{\theta}_{pq}^{(2)}$ are defined as in (\ref{eq:sample_variance_covariance}), and 
\begin{equation}
\hat{\phi}_{pq} = \frac{1}{I} \sum_{I=1}^{n} \left[ \left(\zeta_{ip}^{(1)} - \bar{\zeta}_{p}^{(1)}\right)
\left(\zeta_{iq}^{(1)} - \bar{\zeta}_{q}^{(1)}\right)
\left(\zeta_{ip}^{(2)} - \bar{\zeta}_{p}^{(2)}\right)\left(\zeta_{iq}^{(2)} - \bar{\zeta}_{q}^{(2)}\right) \right] - \hat{\omega}_{pq}^{(1)}\hat{\omega}_{pq}^{(2)}
\end{equation}
captures the covariance between $\hat{\omega}_{pq}^{(1)}$ and $\hat{\omega}_{pq}^{(2)}$. 

We propose a permutation-based approach to obtain the p-value for an observed test statistic using a scheme that generates data under the null while preserving the paired structure. For $1 \leq p \leq P$, construct $\mathbf{\tilde{Y}}_p = \left\{\left[\tilde{Y}{i}^{(1)}(t),\tilde{Y}{i}^{(2)}(t)\right] : i \in \{1, \dots, I\}\right\}$ by randomly permuting $z = \{1,2\}$ within each pair $i$, and let $\tilde{M}_p$ be the corresponding test statistic. A test of level $\alpha$ rejects the null hypothesis of equality of eigenfunctions if $M \geq Q{(1-\alpha)}$, where $Q{(1-\alpha)}$ is the $1-\alpha$ quantile of the empirical distribution of the permuted statistics $\tilde{M}_p$. In practice, we use $P = 1000$ permuted datasets to define the empirical null distribution. 

%%%%%%%%%%%%%%%%%%%%%%%%%%%%%%%%%%%%%%%
\subsection{Practical Implementation}
\label{subsec:implementation}

In real-world applications, functions are observed with noise over finite grids that can be dense or sparse, and regular or irregular. We obtain score estimates $\hat{\boldsymbol{\zeta}}_i^{(z)} = \Big\{\hat{\zeta}_{ik}^{(z)}: k \in \{1,\dots,K\} \Big\}$ of the true scores in pooled model (\ref{eq:null_model}) using existing FPCA methods, and then use these in order to test the hypothesis of equality of eigenfunctions.  

First, we estimate the group-specific means of both groups and remove them; this step is necessary because many implementations of FPCA assume a single mean function, which differs from our framework. For regularly-observed data we suggest using a point-wise estimation of the group-specific means, while for irregularly-observed data we suggest using a smoother-based approach such as $\texttt{gam()}$ in the $\texttt{mgcv}$ package \citep{wood2001mgcv}. Next, we combine de-meaned data from both groups and estimate the shared eigenfunctions, eigenvalues, and subject-specific scores. There are a number of FPCA approaches that could be used and our test is not limited to any specific method. In our simulations and analyses, we use $\texttt{fpca.face()}$ in the $\texttt{refund}$ package \citep{xiao2016fast,goldsmith2016refund}, which is a computationally efficient implementation of FPCA that we have found to be accurate and suitable for the dense, regular grids in our simulations and real data application. Finally, we perform a two-sample covariance test on the covariance of the estimated scores using the function $\texttt{HDtest::testCov()}$ \citep{cao2018package}. Code is publicly available on Github \url{https://github.com/angelgar/funvTests}

Like similar tests for covariance operators, our test requires a fixed number $K$ of FPCs. There is an inherent tradeoff in the choice of $K$: selecting a small value could increase the risk of a type II error if the true difference is in the later eigenfunctions, while a large $K$ could retain unnecessary FPCs and decrease power. Our simulation studies indicate that our test has good power across a wide range of $K$, which reduces sensitivity to this choice. To help mitigate this tradeoff, we suggest using a percent variance explained (PVE) approach, which retains the first $K$ components that explain at least some fixed percentage of the overall variance. We have found that a PVE threshold of 99\% works well in simulations and real data analyses.

%%%%%%%%%%%%%%%%%%%%%%%%%%%%%%%%%%%%%%%
%%%%%%%%%%%%%%%%%%%%%%%%%%%%%%%%%%%%%%%
%%%%%%%%%%%%%%%%%%%%%%%%%%%%%%%%%%%%%%%

\section{Simulations}
\label{sec:sim} 

We present simulation studies designed to assess the empirical size and power of our proposed test across several scenarios. We first conduct simulations where FPCs partially overlap between two groups and then simulations where the FPCs are identical and only eigenvalues differ. Next, we evaluate the properties of our paired test across a range of within-pair correlations. In all cases, we compare the performance of our methodology with two competing methods: the covariance operator test proposed by \citet{panaretos_second-order_2010} and the test for FPC score distributions proposed in \cite{pomann_two-sample_2016}. Furthermore, in the context of paired datasets, we compare both our proposed independent and paired tests to explore the effect of accounting for dependence affects power. We set the type I error rate to be $\alpha = 0.05$ in all tests. Additional simulation results for the somewhat less challenging settings presented in Figure~\ref{fig:method_plot} are given in supporting web materials; those results are qualitatively similar to the ones presented in this Section.

To implement our proposed independent and paired tests, we follow the guidelines in Section \ref{subsec:implementation} and use a PVE threshold of 99\% to select $K$. We implement the test in \citet{panaretos_second-order_2010} using the modified variance-stabilised statistic proposed in their manuscript, which simulation studies suggest it provides higher power. We follow the authors' guidelines to select the optimal number of eigenvectors $K$ using the AIC criterion proposed by \citet{yao2005functional}, and implement it using the R function $\texttt{fdcov::ksample.vstab()}$ \citep{cabassi2016fdcov}. We implement \citet{pomann_two-sample_2016}'s test using R code provided by the authors and select the number of FPCs by retaining the set of FPCs that explain 99\% of the variance. Sensitivity analyses that explore the effect of $K$ on the numerical properties of these tests is available in a supplement.

%%%%%%%%%%%%%%%%%%%%%%%%%%%%%%%%%%%%%%%

\subsection{Simulations for Independent Data}
\label{subsec:sim_ind} 

We generate data from a zero-mean FPCA model in which there are three orthogonal functional principal components:

 \begin{equation}
\label{eq:independent_model}
  Y_i^{(z)}(t) = \sum_{k=1}^{3} \left[\xi_{ik}^{(z)} \frac{ \sin(2\pi kt)}{c_{\phi_k}}\right] + \epsilon_i(t)
\end{equation}

\noindent
where $c_{\phi_k} = {\displaystyle \int_{0}^{1} } \sin(2\pi kt)dt$, $k = \{1,2,3\}$ are normalising constants and the scores $\xi_{ik}^{(z)} $ follow independent univariate Gaussian distributions with mean zero and variance $\lambda^{(z)}_k$. In all simulations $\lambda^{(z)}_1 = 16$ and $\lambda^{(z)}_2=9$ for both $z \in \{1,2\}$. We define $\lambda^{(1)}_3 = \gamma$ and $\lambda^{(2)}_3 = \gamma + \delta$, where $\gamma$ is a baseline shared variance and $\delta$ is an effect size. We generate data for each combination of $\gamma \in \{0,0.1,0.2,0.5,1\}$, $\delta \in \{0,0.1,0.2,0.3,0.4,0.5\}$, creating several scenarios:

\begin{itemize}
\item $\delta = 0, \gamma \geq 0$: null is true;
\item $\delta > 0, \gamma = 0$: null is false; groups have one different FPC;
\item $\delta > 0, \gamma > 0$: null is false; groups share FPCs but have different eigenvalue variances.
\end{itemize}

\noindent
We generate 1000 datasets for each combination of effect size $\delta$, shared variance $\gamma$, and sample sizes $I_1 = I_2 \in \{25,50,100,150,200\}$. Curves are observed over 200 equally spaced points on the grid $\boldsymbol{t} \in [0,1]$. We assume that at each point the data is observed with noise $\epsilon_i(t_{ij})$, which follows a Gaussian distribution with mean zero and variance $\sigma^{2}_\epsilon = 0.25$.

\begin{figure}[ht]
  \centering
     \begin{tabular}{cc}
       \includegraphics[width= 1\textwidth]{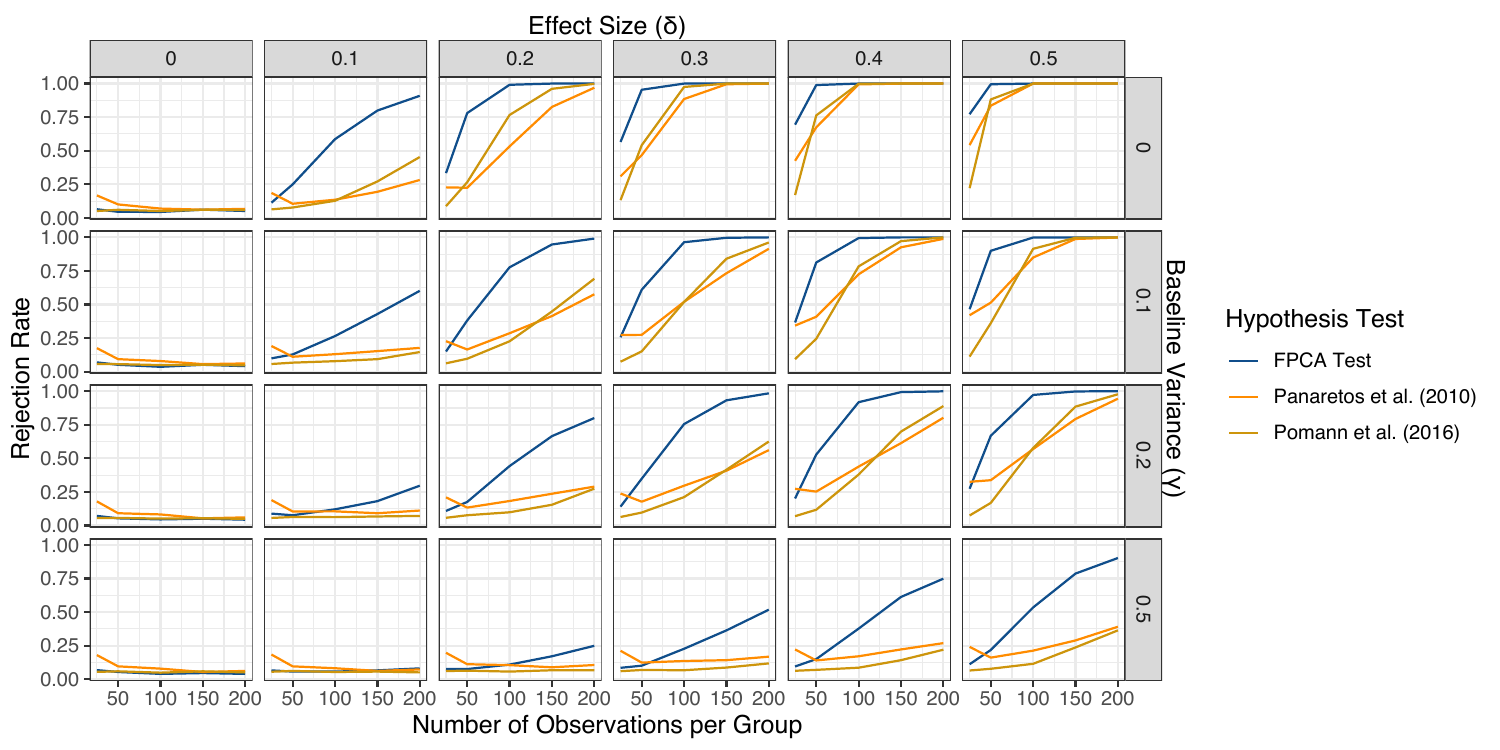} 
     \end{tabular}
     \caption{Empirical rejection rates for independent datasets across simulation settings. We run 1000 simulations for each simulation scenario, and reject the null hypothesis at $\alpha = 0.05$. Our proposed test is in dark blue. Leading competing methods include the test given in \citet{panaretos_second-order_2010} (in orange) and \citet{pomann_two-sample_2016} (in yellow). Each column displays a different effect size, and the rows display the baseline variance shared by both groups where $\lambda_3^{(1)} = \gamma$ and $\lambda_3^{(2)} = \gamma + \delta$}
    \label{fig:sim_independent}
\end{figure}

Results in Figure \ref{fig:sim_independent} show that our proposed test has desirable statistical properties and performs better than competing tests in most situations. Specifically, our test has the correct size and exhibits higher power as the number of observations and the value of  $\delta$ increases.  Comparing our test with existing methods, we found that the test proposed by \citet{panaretos_second-order_2010} tends to reject the null hypothesis too frequently for small sample sizes; meanwhile, for sample sizes where this test has the correct type 1 error rate, it is less powerful than our method. Our results also indicate that our method outperforms the test proposed by \citet{pomann_two-sample_2016} in most scenarios. All methods are most effective in detecting differences in FPCs, and have lower power as the shared variance $\gamma$ increases.

%%%%%%%%%%%%%%%%%%%%%%%%%%%%%%%%%%%%%%%

\subsection{Simulations for Paired Datasets}
\label{subsec:sim_dep} 

We generate synthetic datasets with paired data using a variation of model (\ref{eq:independent_model}). To introduce a paired structure, we simulate scores as correlated multivariate Gaussian variables with mean zero. Specifically, we assume that the $k^{th}$ FPC scores within a pair have correlation $\rho = \text{corr}(\xi_{ik}^{(1)}, \xi_{ik}^{(2)})$, and that this pairwise correlation remains constant across the three FPCs. We fix $\lambda^{(z)}_1 = 16$ and $\lambda^{(z)}_2=9$ for both $z \in \{1,2\}$ and let $\lambda^{(1)}_3 = 0.5$ and $\lambda^{(2)}_3 = 0.5 + \delta$, (i.e. we fix $\gamma = 0.5$) across all simulations.  

To examine the performance of our test under various degrees of dependence, we consider pairwise correlations $\rho \in \{0.2, 0.4, 0.6, 0.8\}$, which is similar to the range of values observed in our motivating data. We estimate empirical size ($\delta = 0$) and  power ($\delta > 0$) through effect sizes $\delta \in \{0, 0.1, 0.2, 0.3, 0.4, 0.5\}$, and choose sample sizes $I_1 = I_2 \in \{25,50,100,150,200\}$. We generate 1000 datasets for each combination, and assume curves are observed over 200 equally spaced points on the grid $\boldsymbol{t} \in [0,1]$. Lastly, We assume that at each point the data is observed with noise $\epsilon_i(t_{ij})$, which follows a Gaussian distribution with mean zero and variance $\sigma^{2}_\epsilon = 0.25$.

\begin{figure}[ht]
  \centering
     \begin{tabular}{cc}
       \includegraphics[width= 1\textwidth]{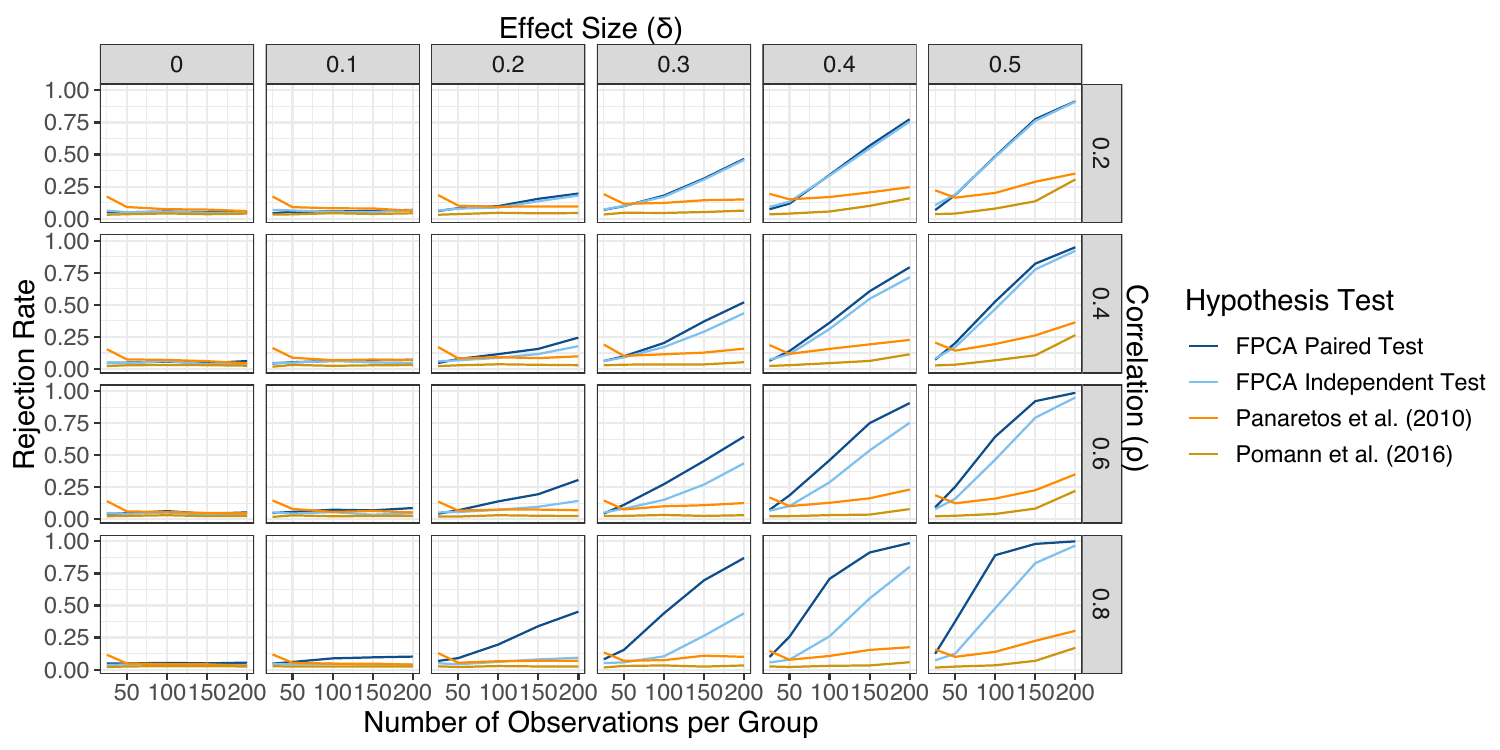} 
     \end{tabular}
     \caption{Empirical rejection rates for paired datasets across simulation settings. We run 1000 simulations for each simulation scenario, and reject the null hypothesis at $\alpha = 0.05$. Our proposed paired test is in dark blue. Competing methods include the proposed independent test is in light blue, and the tests given in \citet{panaretos_second-order_2010} (in orange) and \citet{pomann_two-sample_2016} (in yellow). Each column displays a different effect size, and the rows display the correlation between any two pairs of simulated functions. Across all simulations, $\text{var}(\xi_{i3}^{(1)}) = 0.5$ and $\text{var}(\xi_{i3}^{(1)}) = 0.5 + \delta$}
    \label{fig:sim_paired}
\end{figure}

Figure \ref{fig:sim_paired} indicates that our paired test outperforms competing methods, often substantially, in most scenarios with paired data. Our paired test maintains the correct rejection rate when the null hypothesis is true. Notably, we observe that the performance of our paired test improves as the pairwise correlation increases, which aligns with intuition for paired tests in general. The results for our independent test and both \citet{panaretos_second-order_2010}'s and \citet{pomann_two-sample_2016}'s tests remain similar over the range of pairwise correlations. These tests' performance resembles the independent data results in Figure \ref{fig:sim_independent}.

Sensitivity analyses available in supplementary materials explore scenarios where the number of FPCs, $K$ is common across tests and chosen with three different proportions of variance explained (PVE) thresholds (95\%, 99\%, 99.9\%). The results suggest that our test remains at least equally powerful for independent and paired datasets when compared to competing methods while fixing $K$ to be the same across all methods. Furthermore, we found that both \citet{panaretos_second-order_2010}'s and \citet{pomann_two-sample_2016}'s tests lose power as $K$ increases while our test does not. This is consistent with numerical studies described in \citet{cai_two-sample_2013}, which showed that the power of their test is largely unaffected by the dimension of the covariance matrix. Taken together, these results suggest that our method is robust to reasonable choices of $K$ and has improved performance compared to other methods. 

%%%%%%%%%%%%%%%%%%%%%%%%%%%%%%%%%%%%%%%
%%%%%%%%%%%%%%%%%%%%%%%%%%%%%%%%%%%%%%%
%%%%%%%%%%%%%%%%%%%%%%%%%%%%%%%%%%%%%%%

\section{Motivating Dataset Analysis}
\label{sec:application}

Recall that in the motivating experiment, a trained mouse reached for a food pellet after hearing an auditory cue. Our data consist of activation spike functions observed from 0.25 seconds before the cue to 1.5 seconds after the cue in 25 neurons for each of the 157 trials Figure \ref{fig:raw_data}. Our goal is to test the hypothesis that the latent activation patterns that describe or generate the observed spike functions remain constant across trials.  To visualise trial-to-trial estimates of activation patterns, we begin our analysis by fitting FPCA to each trial using $\texttt{refund::fpca.face()}$; the key question is whether observed differences across trials reflect normal variability in neuron-level activation spikes or can be attributed to true changes in latent activation patterns.

\begin{figure}[ht]
  \centering
     \begin{tabular}{cc}
       \includegraphics[width= 1\textwidth]{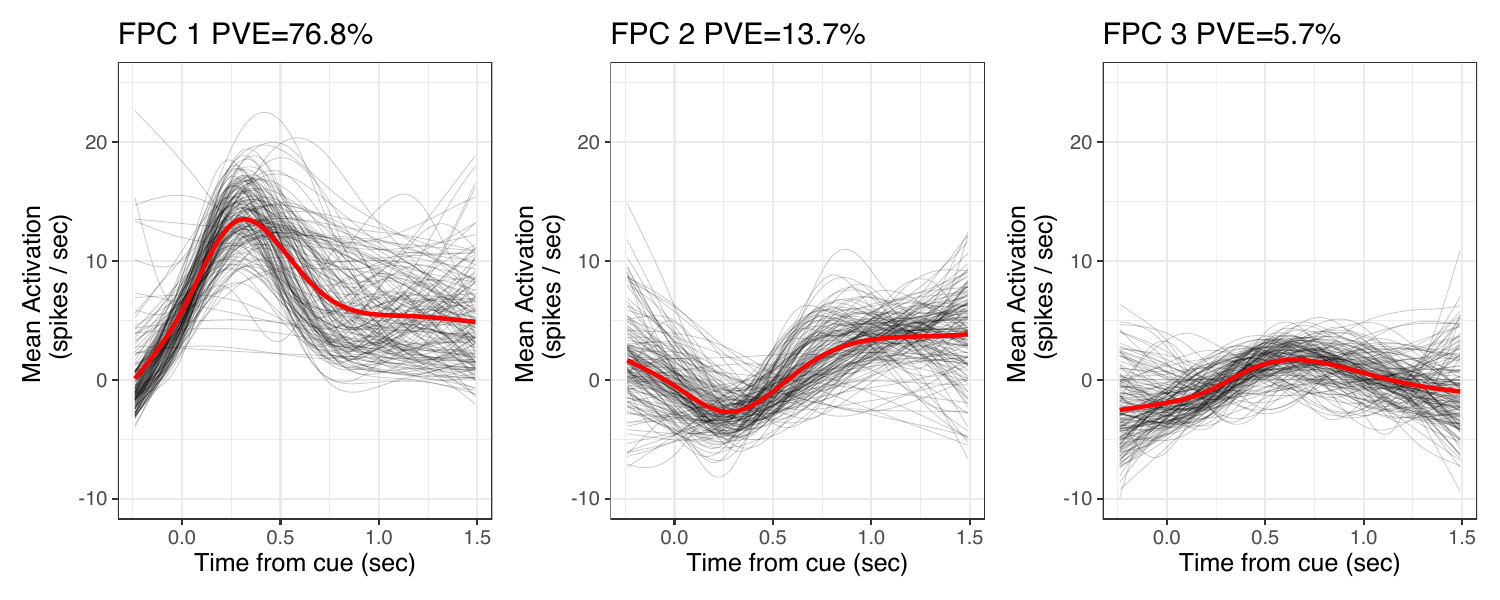} 
     \end{tabular}
     \caption{Spaghetti plots of FPCA decompositions of trial-level data. Each curve represents an estimate for a trial. The panels show the first three FPCs in descending order of most variance explained. On average, these five FPCs explain 96.2\% of the total variability within each trial. The red line is the LOESS average across all trials.}
    \label{fig:fpc_rawdata}
\end{figure}

Figure \ref{fig:fpc_rawdata} displays spaghetti plots of the first three FPCs derived from trial-level decompositions. We show the FPCs in descending order of the amount of variance explained. On average, these 3 FPCs explain 96.2\% of the variability in each trial. The red line is the LOESS mean across all trials for each of the individual components. The first FPC typically explains the intensity of the activation change immediately after the cue and on average explains 76.8\% of the total variance within each trial. The second FPC generally contains a trade-off between initial increased activation and lower baseline activation after the cue; on average, it explains 13.7\% of the total variance within each trial. The third FPC is harder to interpret in general but captures variation unexplained by the first two components; on average, it captures 5.7\% of the observed variability.

\begin{figure}[htbp]
  \centering
     \begin{tabular}{cc}
       \includegraphics[width= 1\textwidth]{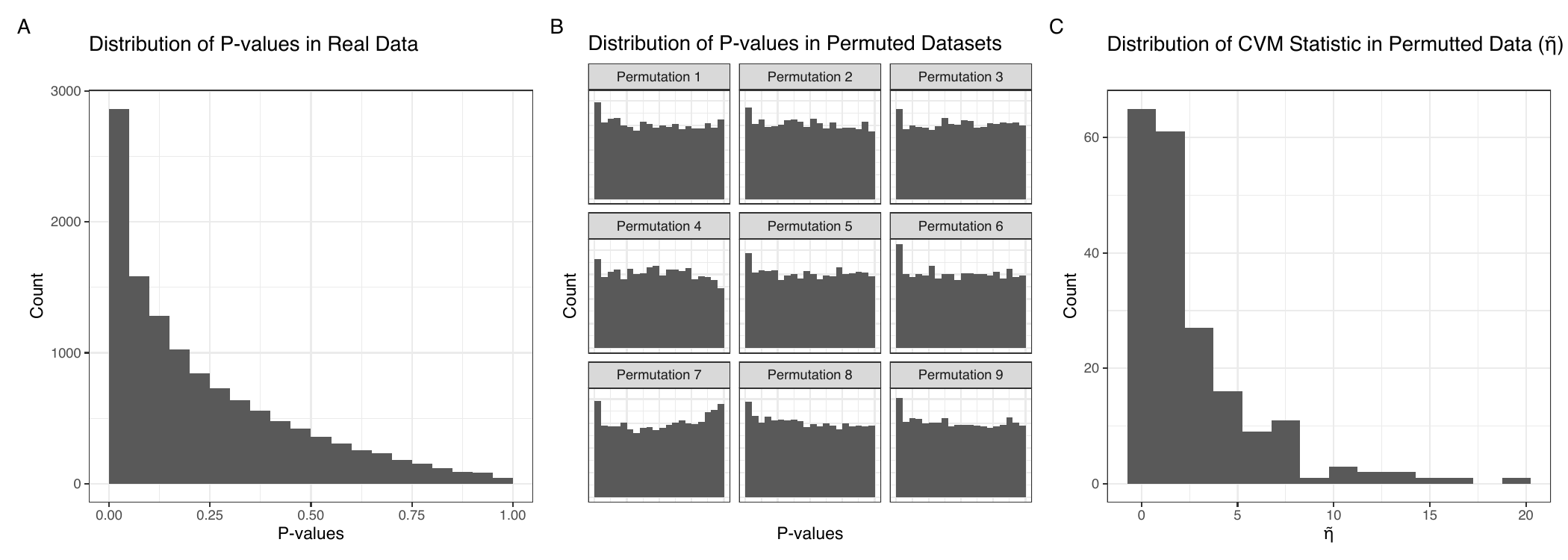} 
     \end{tabular}
     \caption{Panel A displays the distribution of p-values from all pairwise trial comparisons in our motivating dataset. Panel B shows nine example distributions of p-values from all pairwise trial comparisons in permuted datasets in which the null hypothesis is true. Panel C shows the distribution of $\tilde{\eta}_p$ for $p \in \{1,\dots,200\}$.}
    \label{fig:histogram_plot}
\end{figure}

We next apply the test for comparing eigendecompositions for paired data developed in Section \ref{subsec:testing_pairs} to all 12246 pairwise trial comparisons. The resulting distribution $f$ of observed pairwise p-values is shown in the panel A of Figure \ref{fig:histogram_plot}. The distribution is skewed to the right, perhaps indicating lower p-values than expected if all null hypotheses are true. However, since the tests are not independent, we cannot directly compare the distribution $f$ to a known distribution or use these results to test the null hypothesis that the patterns are the same. Instead, we will summarise the p-value distribution with a "global" test statistic and use a permutation approach to assess the null hypothesis that the activation patterns remain constant across trials based on that statistic.

To summarise the distribution $f$, we calculate the Cramer-Von Mises (CVM) statistic by comparing $f$ to the uniform distribution. Let $F$ be the cumulative distribution function of $f$. The CVM statistic is defined as $\eta={\displaystyle \int_{0}^{1} }\left\{\left[F(x)-x\right]^{2}\right\}dx$. We compare the observed value of $\eta$ to an empirical null distribution that we derive using a permutation approach. For each permuted dataset, we begin by permuting the trial labels within each neuron $i$. This scheme preserves the repeated observation structure in the original dataset, as each neuron will only be in a trial once, while removing trial-specific effects and yielding data under the null hypothesis. For each of $P = 200$ permuted datasets, we estimate $\tilde{f}_p$, the distribution of p-values for all pairwise comparisons of trials coming from the $p^{th}$ permuted dataset. We calculate $\tilde{\eta}_p ={\displaystyle \int_{0}^{1} }\left\{\left[\tilde{F}_p(x)-x\right]^{2}\right\}dx$ for $1 \leq p \leq P$, where $\tilde{F}_p$ is the cumulative distribution of $\tilde{f}_p$. Finally, we reject the null hypothesis at a level $\alpha$ if $\eta \geq Q{(1-\alpha)}$, where $Q{(1-\alpha)}$ is the $(1-\alpha)$ quantile of the empirical distribution of the permuted statistics. 

The results of our analysis are shown in Figure \ref{fig:histogram_plot}. Panel B shows nine sample histograms, each of which represents the distribution of all pairwise p-values obtained from the analysis of a single permuted dataset. These are approximately uniform, and much less skewed than the distribution of p-values obtained in the unpermuted dataset. Panel C shows the estimated empirical null distribution of the 200 $\eta$ obtained from the permuted datasets. The range of the empirical distribution is [0.14, 19.52], with a 95\% quantile of 9.98. The test statistic $\eta$ derived using the observed data is 949.14. Therefore, we reject the null hypothesis at a significance level of $\alpha = 0.05$, concluding that activation patterns differ across trials in a manner not attributable to neuron-level noise.

\begin{figure}[htb]
  \centering
     \begin{tabular}{cc}
       \includegraphics[width= 1\textwidth]{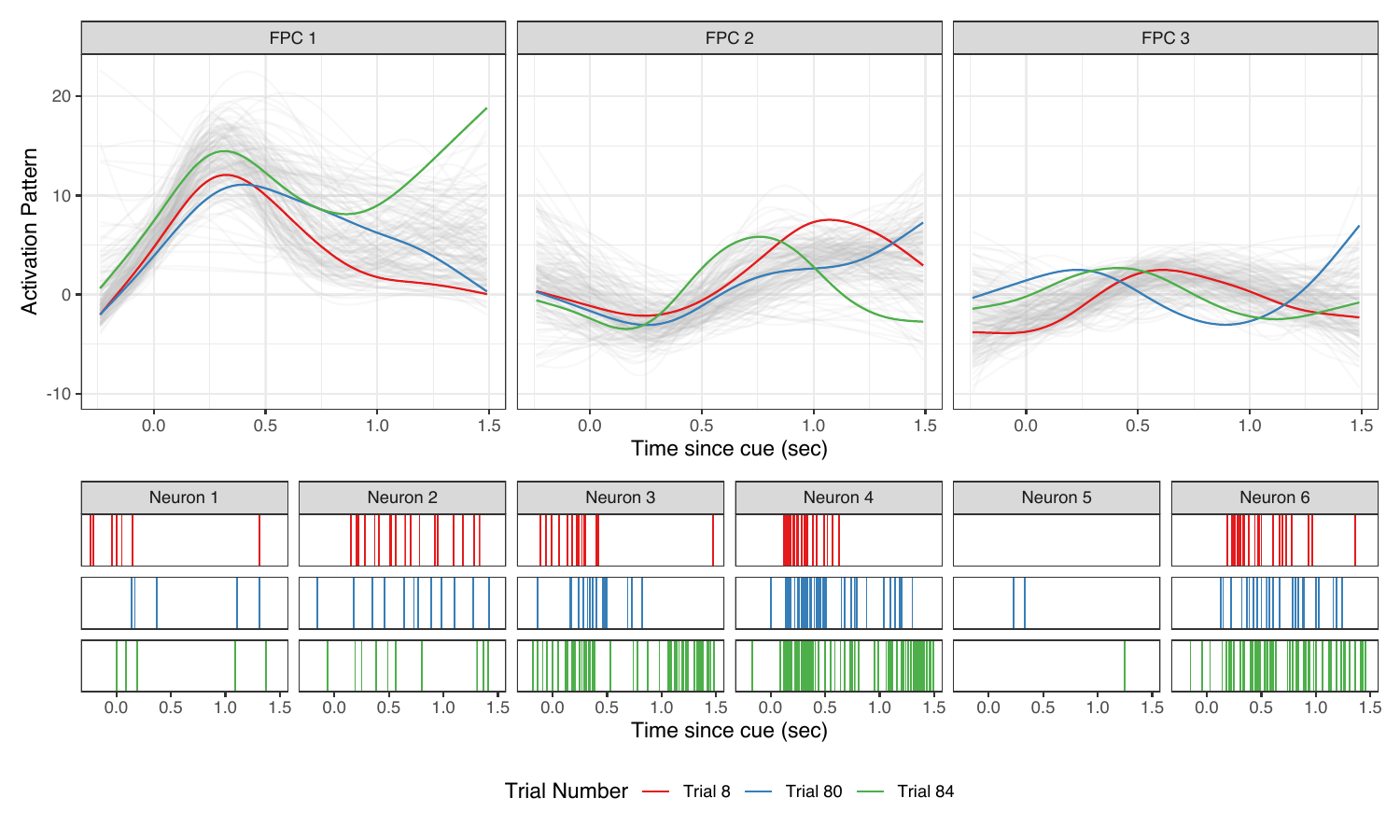}
     \end{tabular}
     \caption{The top panel highlights activation patterns for three example trials. We observe marked differences in the later stages of the observation window in the first pattern for the three highlighted trials. Bottom panel shows barcode plots of raw dichotomised neural spike data. Each coloured line represents a timepoint in which that neuron was active. There are differences in raw activation across the three trials, as evidenced by the activation patterns.}
    \label{fig:two_comparisons}
\end{figure}

Although our global test does not provide information about the significance of individual comparisons or identify specific trials that exhibit greater than expected differences, we can provide exploratory evidence to support the conclusion that latent activation patterns differ across trials. Figure \ref{fig:two_comparisons} presents an overview of the activation patterns and neural activity in three illustrative trials. The top panel highlights trial-specific activation patterns obtained through FPCA for these three trials (FPCA results for all trials are shown in the background). The first FPC for Trial 84 exhibits a late activation across neurons, in contrast to the return to baseline activation observed in Trials 8 and 80. The bottom panel of Figure \ref{fig:two_comparisons} shows the raw activation for these three trials in the neurons shown in Figure \ref{fig:raw_data}. The firing spikes for these trials is consistent with the activation patterns identified through FPCA: the neurons' behaviour is similar across all trials at the beginning of the task but diverge roughly 1 second after the auditory cue. Specifically, neurons 3, 4, and 6 exhibit increased activity in Trial 84 compared to Trials 8 and 80. Although pairwise p-values are not adjusted for multiple comparisons, our test suggests that activation patterns in Trial 84 differ from those in Trial 8 ($p < 0.01$) and those in Trial 80 ($p < 0.01$). There is no evidence that activation patterns differ between Trial 8 and Trial 84 ($p = 0.72$), suggesting that differences in estimated patterns reflect only trial-to-trial noise in neural firing.

%%%%%%%%%%%%%%%%%%%%%%%%%%%%%%%%%%%%%%%
%%%%%%%%%%%%%%%%%%%%%%%%%%%%%%%%%%%%%%%
%%%%%%%%%%%%%%%%%%%%%%%%%%%%%%%%%%%%%%%

\section{Discussion}
\label{sec:disc}

We have developed a novel method for testing whether two sets of functions share a common eigendecomposition by analysing the covariance matrix of the scores obtained from an FPCA decomposition. Our approach includes two versions of the test, one for independent functions and the other for pairs of functions, and we demonstrate that our method is more powerful than current competing tests while retaining the correct empirical size under the null hypothesis.

We sought to investigate whether neural spike data from a mouse trained to reach for a pellet displayed trial-to-trial variability in its activation patterns. To test this hypothesis, we performed all pairwise comparisons of trials and used the distribution of p-values to summarise our findings. We compared this distribution to the null distribution by utilising a permutation-based approach. Our results suggest the presence of trial-to-trial variability in the activation patterns, and may indicating that future analyses of neural spike data should account for and leverage this source of variation. For instance, changes in neural activation patterns may be related to observable behavioural differences, including variability in reaches across trials. 

Finally, we note some limitations of our method and directions for future work. We focus exclusively on the covariance structures in two samples, and will be unable (by design) to detect differences in the mean. More subtly, our test is not sensitive to changes in score distributions that do not affect the variance: if two samples have the same FPCs and same score variance but scores are unimodal in one sample and bimodal in the other, our test will fail. We therefore encourage careful inspection of score distributions, and note that \citet{pomann_two-sample_2016} may be better suited to this setting. An extension of our method could analyse our motivating data as dichotomous or counts, rather than assuming Gaussian distributions, using techniques in Generalised FPCA \citep{goldsmith2015generalized,gertheiss2017note}. We use estimated scores to construct the sample covariance matrices that are necessary for our test. In our setting scores are well-estimated for both groups, but additional work may be needed when variance in score estimates is large or dissimilar across groups due to study design (e.g. because one group is measured on dense grid and the other is measured on a sparse grid). Finally, investigating how differences in latent activation patterns across trials is related to differences in the produced reaching trajectories is an subject of future work to understand skilled movement. 

%%%%%%%%%%%%%%%%%%%%%%%%%%%%%%%%%%%%%%%
%%%%%%%%%%%%%%%%%%%%%%%%%%%%%%%%%%%%%%%
%%%%%%%%%%%%%%%%%%%%%%%%%%%%%%%%%%%%%%%

\section{Acknowledgments}
\label{sec:ack}

This work was supported by Award R01NS097423-01 from the National Institute of Neurological Disorders and Stroke.

%%%%%%%%%%%%%%%%%%%%%%%%%%%%%%%%%%%%%%%
%%%%%%%%%%%%%%%%%%%%%%%%%%%%%%%%%%%%%%%
%%%%%%%%%%%%%%%%%%%%%%%%%%%%%%%%%%%%%%%

\newpage
\bibliographystyle{rss}
\bibliography{biblio}

%%%%%%%%%%%%%%%%%%%%%%%%%%%%%%%%%%%%%%%
%%%%%%%%%%%%%%%%%%%%%%%%%%%%%%%%%%%%%%%
%%%%%%%%%%%%%%%%%%%%%%%%%%%%%%%%%%%%%%%

\newpage

\section{Supporting Web Documents}
\beginsupplement

\subsection{Analysis of Simulations with Different Eigenfunctions}
\label{subsec:supp_sim1} 

We first describe the data-generating models used in the two scenarios shown in Figure \ref{fig:method_plot} to illustrate our methodology in Section \ref{subsec:testing_procedure}. We then proceed to present a power analysis of our proposed methodology under these scenarios. 

We generate data $\left[Y_{i}^{(z)}(t) : i \in \left\{1, \dots, I_{z}\right\}\right]$ for group $z \in \{1,2,3\}$ using zero-mean FPCA models:
\begin{equation} 
\label{eq:eigenfunc_sim_models}
\begin{split}
Y_i^{(1)}(t) &= \sum_{k=1}^{2} \left[\xi_{ik}^{(1)} \frac{ \sin(2\pi kt)}{c_{\phi_k}^{(1)}}\right] + \epsilon_i(t) \\
  Y_i^{(2)}(t) &= \sum_{k=1}^{2} \left[\xi_{ik}^{(2)} \frac{ \cos(2\pi kt)}{c_{\phi_k}^{(2)}}\right] + \epsilon_i(t) \\
  Y_i^{(3)}(t) &= \sum_{k=1}^{2} \left[\xi_{ik}^{(3)} p_k(t) \right] + \epsilon_i(t) \\
\end{split}
\end{equation}

\noindent
where $c_{\phi_k}^{(1)} = \int_{0}^{1} \sin(2\pi kt)dt$ and $c_{\phi_k}^{(2)} = \int_{0}^{1} \cos(2\pi kt)dt$, $k \in \{1,2\}$ are normalising constants, and $p_k(t), k \in \{1,2\}$ are orthonormal polynomials as implemented in the R function \texttt{stats::poly()}. The scores $\xi_{ik}^{(z)}, z \in \{1,2,3\}$ follow independent univariate Gaussian distributions with mean zero and variance $\lambda^{(z)}_1 = 4$ and $\lambda^{(z)}_2= 1$ for $z \in \{1,2,3\}$. We assume that at each point the data is observed with noise $\epsilon_i(t_{ij})$, which follows a Gaussian distribution with mean zero and variance $\sigma^{2}$.

We compare three scenarios: one where the null hypothesis is true by generating two datasets from model $z = 1$, another where the eigenfunctions across both groups are orthogonal using data from $z = 1$ vs. $z = 2$ (Panel A1-4 in Figure \ref{fig:method_plot}), and a third scenario where the eigenfunctions across groups are not orthogonal, using data from $z = 1$ vs. $z = 3$ (Panel B1-4 in Figure \ref{fig:method_plot}). We generate 1000 datasets for each combination of sample sizes $I_1 = I_2 \in \{25, 50, 100, 150, 200\}$ and error variances $\sigma^{2}_\epsilon \in \{0.25, 0.5, 1\}$. The curves are observed over 200 equally spaced points on the grid $\boldsymbol{t} \in [0,1]$. For each dataset, we implement the three methods previously described in Section \ref{subsec:sim_ind} and reject the null hypothesis at $\alpha = 0.05$ and summarise the rejection rate across the 1000 datasets. All three tests were implemented as described in Section \ref{subsec:sim_ind}.

\begin{figure}[htb]
  \centering
     \begin{tabular}{cc}
       \includegraphics[width= 0.7\textwidth]{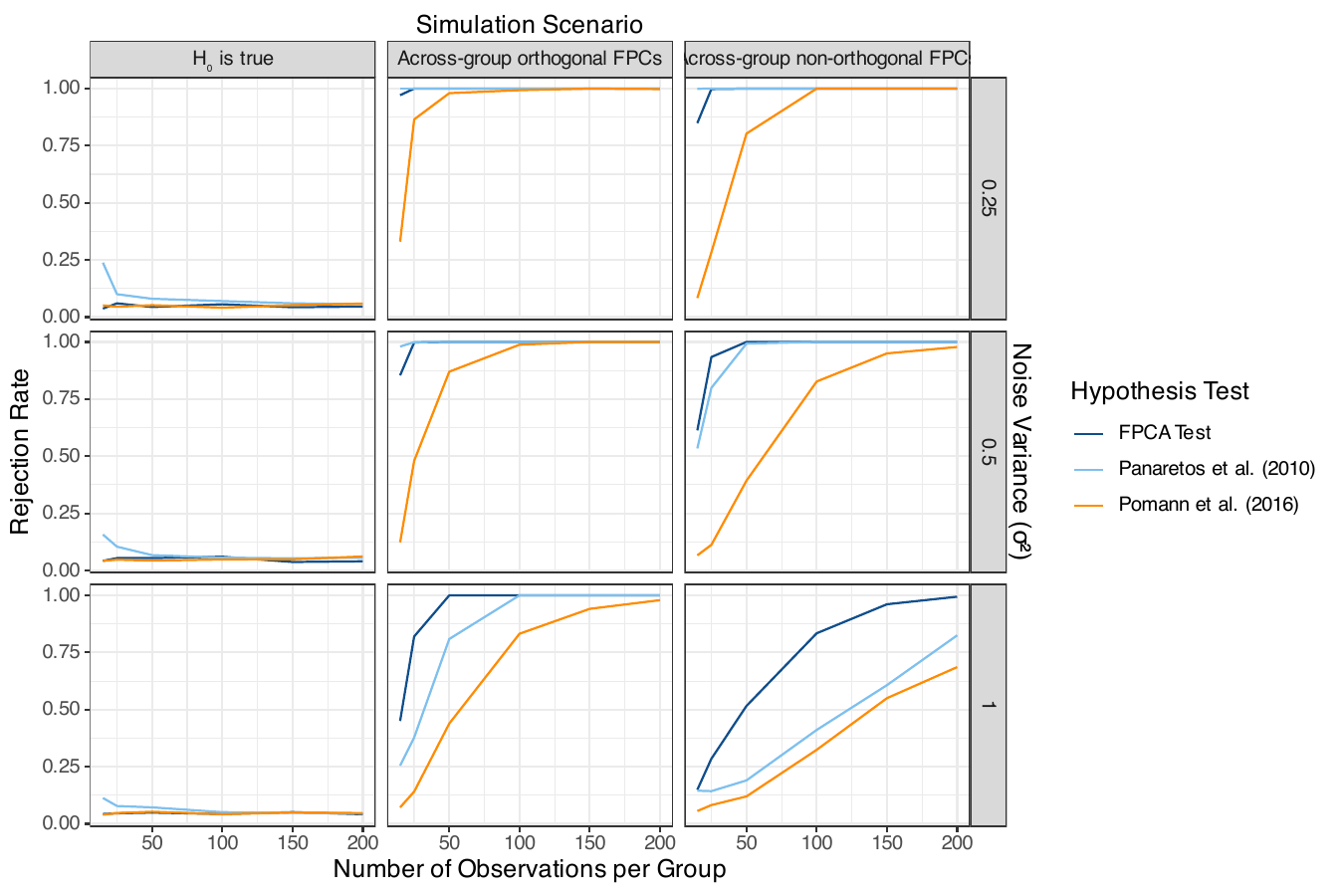} 
     \end{tabular}
     \caption{Empirical rejection rates based on sample size are presented from 1000 simulations for each scenario. The null hypothesis was rejected at $\alpha = 0.05$. Our proposed test is indicated in dark blue, while leading competing methods from \citet{panaretos_second-order_2010} (in orange) and \citet{pomann_two-sample_2016} (in yellow) are included. The left column displays an scenario in which $H_0$ is true. The middle column displays FPCs orthogonal across-groups, and the right column shows non-orthogonal FPC scenarios. Rows represent different error variances $\sigma^2$.}
    \label{fig:supp_twoscenarios}
\end{figure}

The results in Figure \ref{fig:supp_twoscenarios} demonstrate that our proposed test has favourable statistical properties. Specifically, it has the correct size and outperforms competing tests in both scenarios. When compared to existing methods, the test by \citet{panaretos_second-order_2010} tends to over-reject the null hypothesis for small sample sizes, while being less powerful than our method for sample sizes with correct type 1 error rates. All tests show reduced power as the error variance $\sigma^2$ increases and higher power as the number of observations increases. Additionally, all methods perform better at detecting differences when the FPCs across the two groups are orthogonal. These tests rely on pooled orthogonal eigenfunctions. We hypothesise that when the eigenfunctions are orthogonal across groups, the pooled eigenfunctions closely approximate true eigenfunctions, leading to better-score estimation and a more powerful test. Furthermore, \citet{pomann_two-sample_2016}'s test relies on univariate comparisons of score distributions and ignores any correlation that arises in non-orthogonal scenarios.

\subsection{Sensitivity Analysis of \textit{K} on proposed Methodology}
\label{subsec:supp_sim_k} 

We present sensitivity analysis to explore the role that the number of functional principal components (FPCs) plays in the numerical properties of our proposed methodology. Our goal is to compare the methods when fixing $K$, and to examine the impact that $K$ has on the empirical rejection rates. To do this, we generate synthetic datasets for two groups of independent functional data as described in Section \ref{subsec:sim_ind}. For each dataset, we estimate $K$ using three different proportion of variance explained (PVE) thresholds (95\%, 99\%, 99.9\%) and implement the three methods previously described in Section \ref{subsec:sim_ind} with the $K$ values selected from these three criteria. 

\begin{figure}[htb]
  \centering
     \begin{tabular}{cc}
       \includegraphics[width= 0.8\textwidth]{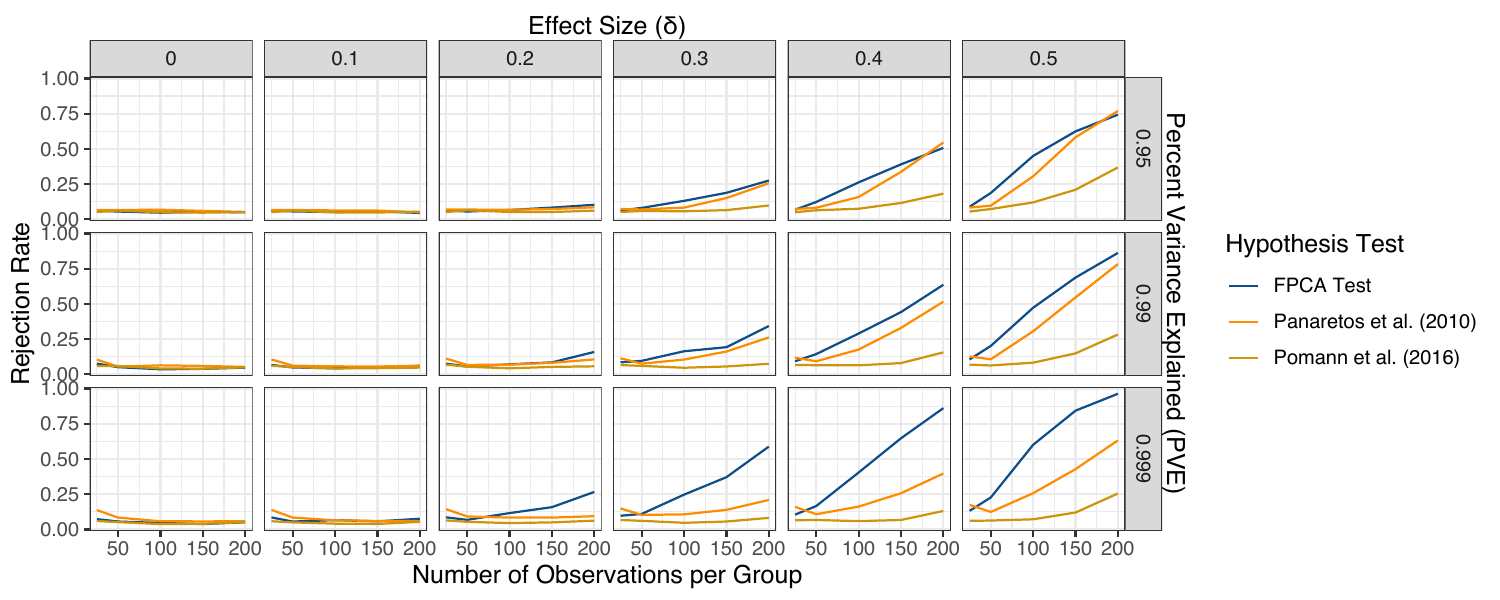} 
     \end{tabular}
     \caption{Plot of Empirical Rejection Rates for Independent Data when PVE = (95\%, 99\%, 99.9\%). We run 1000 simulations for each scenario and reject the null hypothesis at a 5\% level. Our proposed test for independent data is in dark blue, \citet{panaretos_second-order_2010}'s test is in orange and  \citet{pomann_two-sample_2016}'s test is in yellow. We present the scenario $\gamma = 0.5$}
    \label{fig:supp_independent}
\end{figure}

Figure \ref{fig:supp_independent} summarises the simulation rejection rates for the three PVE thresholds (95\%, 99\%, 99.9\%). Our proposed method exhibits higher power when a 99.9\% threshold is used. In contrast, we observe that both \citet{panaretos_second-order_2010} and \citet{pomann_two-sample_2016}'s methods perform better for smaller values of $K$. Our findings suggest that our methodology is robust to reasonable choices of $K$, as our method's power remains relatively constant for larger values of $K$, while competing methods lose power. Our method is at least equally powerful as competing methodologies when fixing the value of $K$.

\begin{figure}[ht]
  \centering
     \begin{tabular}{cc}
       \includegraphics[width= 0.8\textwidth]{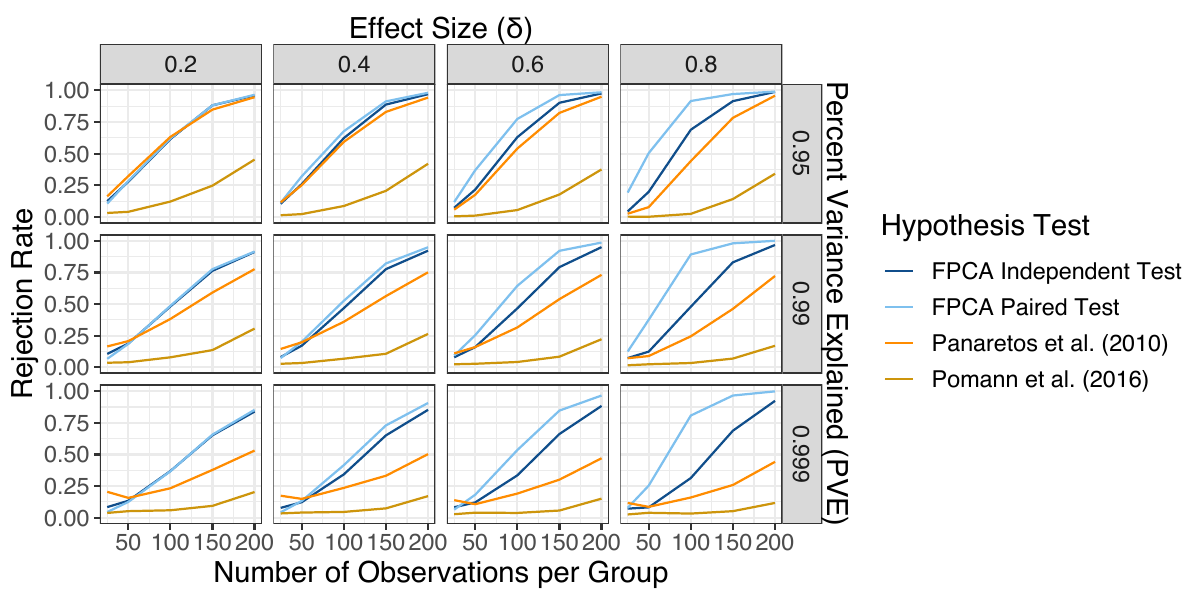} 
     \end{tabular}
     \caption{Plot of Empirical Rejection Rates for Paired Data when PVE = 99\%. We run 1000 simulations for each scenario and reject the null hypothesis at a 5\% level. We use the percent variance explained criterion and only test the first FPCs that explain 95\% of the variance. Our proposed paired data test is in dark blue, and the proposed independent test is in light blue. The test in  \citet{panaretos_second-order_2010} is in orange and the test in \citet{pomann_two-sample_2016} is in yellow. We present the scenario $\gamma = 0.5$ and $\delta = 0.5$}
    \label{fig:supp_dependent}
\end{figure}

We now explore the paired data scenario by generating synthetic datasets for two groups of paired functions as described in Section \ref{subsec:sim_dep}. For each dataset, we estimate K using three PVE thresholds (95\%, 99\%, 99.9\%) and implement the four methods previously described in that same section. Figures \ref{fig:supp_dependent} summarise the simulation rejection rates for the three PVE thresholds used (95\%, 99\%, 99.9\%). The results are similar to Figure \ref{fig:supp_independent}. We found that our test is somehow stable across values of PVE and that both \citet{panaretos_second-order_2010} and \citet{pomann_two-sample_2016}'s tests perform better for smaller values of $K$. Our method is at least equally powerful as competing methodologies when fixing the value of $K$.

%%%%%%%%%%%%%%%%%%%%%%%%%%%%%%%%%%%%%%%
%%%%%%%%%%%%%%%%%%%%%%%%%%%%%%%%%%%%%%%
%%%%%%%%%%%%%%%%%%%%%%%%%%%%%%%%%%%%%%%
%%%%%%%%%%%%%%%%%%%%%%%%%%%%%%%%%%%%%%%

\end{document}